\documentclass{iau}
\usepackage{graphicx}
\usepackage{natbib}
\usepackage{color}
\definecolor{darkgreen}{RGB}{0,142,128}

\graphicspath{{Figs/}}

\title[AS IAUS300~ Modeling SPMIs: boundary condition effects]{Modeling magnetized star-planet
  interactions: boundary conditions effects}

\author[Antoine Strugarek et al.]{Antoine Strugarek$^{1,2}$ \and Allan
  Sacha Brun$^2$ \and Sean P. Matt$^3$ \and Victor Reville$^2$}

\affiliation{
$^1$D\'epartement de physique, Universit\'e de
  Montr\'eal, \\
 C.P. 6128 Succ. Centre-Ville, Montr\'eal, QC H3C-3J7, Canada \\
 email: {\tt strugarek@astro.umontreal.ca} \\[\affilskip]
$^2$Laboratoire AIM Paris-Saclay, CEA/Irfu Universit\'e Paris-Diderot
CNRS/INSU, \\
 F-91191 Gif-sur-Yvette \\[\affilskip]
$^3$Department of Physics \& Astronomy, University of Exeter, Exeter EX2 4QL, UK}

\pubyear{2013}
\volume{300}  %% insert here IAU Symposium No.
\pagerange{xxx--xxx}
\jname{Nature of prominences and their role in Space Weather}
\editors{A.C. Editor, B.D. Editor \& C.E. Editor, eds.}
\begin{document}

\maketitle

\begin{abstract}
We model the magnetized interaction between a star and a close-in
planet (SPMIs), using global, magnetohydrodynamic numerical
simulations.  In this proceedings, we study the effects of the
numerical boundary conditions at the stellar surface, where the
stellar wind is driven, and in the planetary interior.  We show that
is it possible to design boundary conditions that are adequate to
obtain physically realistic, steady-state solutions for cases with
both magnetized and unmagnetized planets. This encourages further
development of numerical studies, in order to better constrain and
undersand SPMIs, as well as their effects on the star-planet
rotational evolution. 
\end{abstract}

\firstsection % if your document starts with a section,
\section{Introduction}

The growing number of known exoplanet systems raises the question of
how to properly define the habitability zone around a star \citep{Kasting:1993hw,Barnes:2011tc}. Its
definition depends on the interactions existing
between a planet and its host star, which are gravitational (tidal forces),
magnetic (wind-planet interactions, hereafter referred to as SPMI) and
radiative (\textit{e.g.}, stellar EUV ionisation flux).
Magnetized interactions between a star and its orbiting planets have
recently been suggested to be at the origin of a possibly enhanced planet detectability
\citep{Jardine:2008ec,Fares:2010hq,Miller:2012gq}. In the case of a
close-in planet, these interactions may also be at the origin of anomalous stellar
magnetic activity \citep{Cuntz:2000ef,Lanza:2008fn,Donati:2008hw}. It
was also suggested that it could affect the star-planet rotational
evolution
\citep{Laine:2008dx,Pont:2009ip,Cohen:2010jm,Vidotto:2010iv,Lanza:2010bo}. Theoretical 
work is needed to better understand SPMIs.

Based on a pioneering work done in the context of the satellites of Jupiter
\citep{Goldreich:1969kf,Kivelson:2004vf}, \citet{Laine:2011jt} built
an analytical model describing the various
components of SPMIs in the case of unmagnetized
planets. Pursuing the same goal, \citet{Lanza:2013gj} also developed
semi-analytical models of SPMIs in the context of magnetized
planets. However, a systematic numerical validation of those models still
remains to be properly done \citep[see][for first
steps towards such a validation]{Ip:2004ba,Cohen:2011gg}. 

Focusing on close-in planets, the SPMIs include magnetic reconnection,
magnetic field  diffusion at
the stellar surface and in the planet vicinity,
radiation and ionisation processes in the planetary magnetosphere and
magneto-sonic wave propagation.
A numerical investigation of SPMI requires a careful description
of those physical processes although it is generally not possible to
treat all of them simultaneously with a unique model. Hence,
specific strategies such as dedicated boundary conditions have to be
developed to study SPMI from a global point of view. We detail in this
work how to develop both stellar (section \ref{sec:star-bound-cond})
and planetary (section \ref{sec:plan-bound-cond}) boundary conditions
to globally model the different SPMI cases,
within the MHD formalism.

\section{Stellar boundary conditions}
\label{sec:star-bound-cond}

\begin{figure}[b]
\begin{center}
  \includegraphics[width=0.9\linewidth]{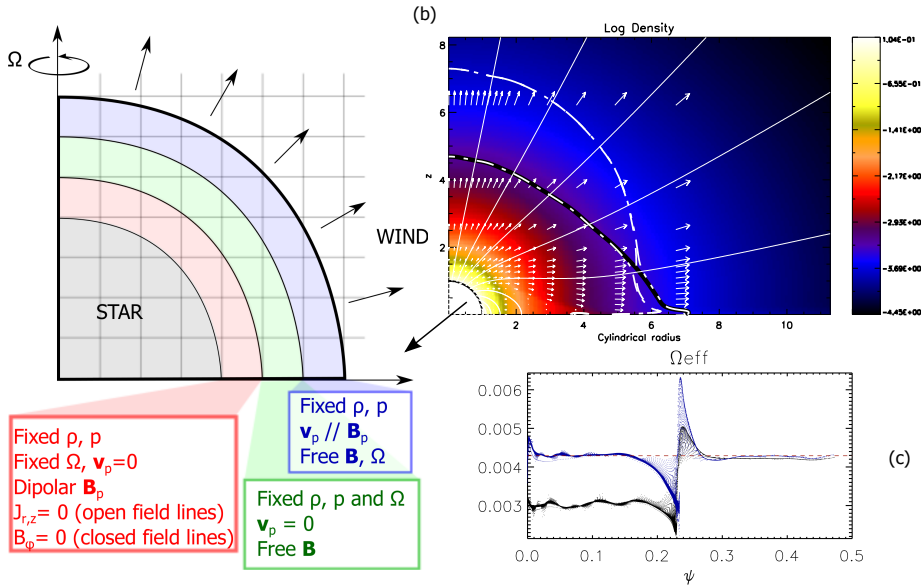} 
 \caption{\textbf{(a)} Schematic of the multi-layer boundary condition
 ensuring good conservation properties of the MHD solution as well as
 reactivity to external stimuli. Fixed quantities are forced to the
 Parker wind solution. The subscript $p$ stands for the poloidal
 component $(\varpi,z)$ of vector in cylindrical
 coordinates. \textbf{(b)} Typical wind solution 
 used for SPMI. The color map represents the logarithmic density, the white lines
 the poloidal magnetic field lines. The slow and fast Alfv\`en surfaces are labeled
 by the dashed lines, and the arrows show velocity field. The stellar
 surface is labeled by a black quarter of a circle. The axes are in
 stellar radius units. \textbf{(c)}
 Effective rotation rate as a function of the streamfunction
 $\psi$ for good (blue dots) and bad (black dots) boundary conditions. The red dashed horizontal line labels the stellar rotation
 rate. Low values of $\psi$ correspond to open polar field lines and
 larger values of $\psi$ to closed equatorial field lines. Each dot corresponds to a grid point.}
   \label{fig:fig1}
\end{center}
\end{figure}

We model stellar winds following numerous previous
analytical and numerical studies
\citep{Weber:1967kx,Washimi:1993vm,Ustyugova:1999ig,Keppens:2000ea,Matt:2004kd,Matt:2012ib}. We use standard
MHD theory to numerically model with the PLUTO code
\citep{Mignone:2007iw} magnetized steady state flows 
anchored at the surface 
of a rotating star. We model winds driven by the thermal pressure of
the stellar corona in a 2D axisymmetric cylindrical geometry \citep[see][for a more detailed description of the MHD
model we use]{Strugarek:2012th}. 

The steady-state wind solution can depend very sensitively on the
type of boundary conditions that are imposed under the stellar
surface. Because we want to use our model to study SPMIs, the stellar boundary
conditions have to be able to both react and adapt to external stimuli
originating from the orbiting planet. The design of a boundary
condition satisfying those two conditions, and its associated stellar
wind solution, are displayed in panels (a) and (b) of fig. \ref{fig:fig1}.

We developed a layered boundary condition over which the stellar wind
characteristics are progressively enforced as we go deeper
under the stellar surface. This boundary condition ensures 
very good conservation properties \citep{Lovelace:1986kd,Zanni:2009kc}
along the magnetic field lines. This 
is exemplified in panel (c) of fig. \ref{fig:fig1}. We display 
the effective rotation rate $\Omega_{\mbox{eff}} \equiv \frac{1}{\varpi}\left(v_\phi
  -\frac{v_p}{B_p}B_\phi \right)$ as a function of the streamfunction
$\psi$ generating the poloidal magnetic field. In a steady-state,
ideal MHD wind, $\Omega_{\mbox{eff}}$ should be constant along each
field line and equal to $\Omega_{\star}$. The blue dots
correspond to the boundary condition described in panel (a), and the
black dots to a case where $B_{\phi}$ is set to 0 at all latitudes in
the third boundary level. We observe that the target
stellar rotation rate (dashed horizontal red line) is
recovered only with the correct boundary conditions. Conservation errors
exist at the open-closed field lines
boundary ($\psi \sim 0.23$), but they remain confined to very few grid
points in the simulation domain. Finally, this boundary condition is intrinsically
able to react to a perturbation by a planet orbiting a star by,
\textit{e.g.}, modifying the stellar wind topology.
We discuss now the importance of
planetary boundary conditions when studying SPMIs.

\section{Planetary boundary conditions}
\label{sec:plan-bound-cond}

\begin{figure}[b]
\begin{center}
  \includegraphics[width=0.9\linewidth]{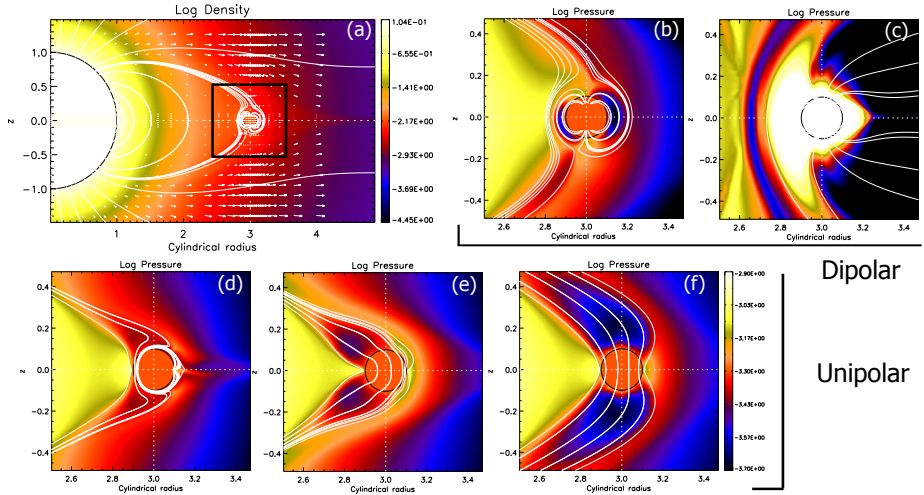}
 \caption{Zoom on planetary boundary conditions effects for dipolar (upper panels) and
   unipolar (lower panels) interactions. The color map represents the
   gas pressure in logarithmic scale,
   and the white lines the magnetic field lines. The planet surface is
   labeled by a black circle at $1$ stellar radius. Panels (a) and (b) show the fiducial dipolar case,
 and panel (c) is the unrealistic case of a planet with a very high
 internal pressure. Panel (d) represents a
 Venus-like interaction and panels (e) and (f) two Io-Jupiter-like interactions.}
   \label{fig:fig2}
\end{center}
\end{figure}

SPMIs are generally decomposed in two categories: the so-called
unipolar and dipolar interactions \citep{Zarka:2007fo}, which refer to 
the cases of unmagnetized and magnetized planets.
Both interactions can be modeled
within the MHD formalism with an adequate boundary condition design. We
detail in this section how to design such boundary conditions. The
examples given here were all done for a planet with a radius of $r_p = 0.1\,
r_\star$, a mass of $M_p=0.01\, M_\star$, an orbital radius
of $r_{\rm{orb}}=3r_\star$ and a resolution of $0.03\, r_p$ at the
planetary surface.

We consider the planet itself as a boundary condition. The PLUTO
code allows one to define internal domains as boundary conditions
over which all variables can be altered during the model evolution. In
all cases, we set the poloidal velocity to zero and the
azimuthal velocity to the keplerian velocity inside the planet. We
also set the density and pressure values inside the planet to 
fiducial values which are consistent with its gravity field. These
value have to be carefully prescribed since they can trigger undesirable
effects in the vicinity of the planet. We give an example of a
dipolar case in panels (a) and (b) of fig. \ref{fig:fig2} (the planetary
magnetic field is simply enforced in the planetary interior in this
case). A stable configuration is obtained when the magnetic pressure and the
gas pressure equilibrate at the interface between the planetary
magnetosphere and the stellar wind. The ram pressure plays little role
here because the planet we consider is in the so called
\textit{dead-zone} of the stellar wind, in which the poloidal
velocity is negligible. We show in panel (c) the exact same simulation for
an extreme case where we multiplied the internal pressure of the
planet by a factor of $20$. The former pressure balance then fails and
a wind is driven from the planet itself. The planetary dipole opens up and a shock
eventually creates at the interface between the two
``winds''. Such undesirable effects may also be obtained by varying the
density of the interior of the planet. Hence, any SPMI model must be developed
to minimize such undesirable effects in the final solution.

Modeling a planet in the unipolar case is a bit more complex than in the
dipolar case. Two classes of unipolar interactions can indeed be
distinguished: Venus-like interaction (case V) and Io-Jupiter like
interaction (case IJ). Note however that in both cases, we consider a
planet located inside the stellar wind dead-zone, at $r_{p}=3\, R_{\star}$.

In case V, the ionisation of the planetary atmosphere 
by the stellar EUV radiation flux allows the creation of a
ionosphere which  acts
as a barrier between the stellar wind magnetic field and the 
unmagnetized interior of the planet \citep{Russell:1993jk}. Depending
on the stellar wind conditions around the planet, an induced
magnetosphere may then be sustained on secular time scales. We show
case V in panel (d) of fig. \ref{fig:fig2}. The
ionosphere is modeled as a very thin ($< 0.2\,
r_{p}$) highly conductive boundary layer under the 
planetary surface. The wrapping of the magnetic field lines around
planet \citep{Russell:1993jk} is naturally recovered.

In case IJ, no ionosphere is created and the stellar wind magnetic field
pervades inside the planet. The SPMI then depends on the ratio of
electrical conductivities between the planetary interior and the stellar
surface where the magnetic field lines are \textit{a priori}
anchored. This ratio sets the effective drag the planet is able to
induce on the stellar wind magnetic field lines. We use the ability of
the PLUTO code to add extra ohmic diffusion in
the planet interior to model it and show in
figure \ref{fig:fig2} two extreme cases in which
magnetic field lines are dragged (panel e) or diffused (panel f) by
the planet. In all cases, we obtain a statistical steady state in
which the SPMI can be analyzed in details. 

\section{Conclusions}
\label{sec:conclusions}

We showed in this work that is it possible to model the global,
magnetized and non-linear interactions between a star and
a planet, within the MHD formalism. It requires a careful
development of adequate boundary conditions to represent
the various interaction cases. We showed that boundary conditions play
a very important role both at the stellar surface and in the planetary
interior. Steady state solutions could be found in the dipolar case as
well as in both the Venus-like and Io-Jupiter-like unipolar
cases. 

The SPMI model we developed will be useful for exploring
stable interaction configurations between a close-in planet and its
host star. In addition, it will enable quantitative predictions of
rotational evolution of star-planet systems due to the
effective magnetic torques which develop in the context of dipolar and unipolar
interactions \citep{Strugarek:2013uh}. Finally, such models
could also be used to study potential SPMI induced emissions, which we
will analyze in a future work.

\acknowledgements

We thank A. Mignone and his team for making the PLUTO code
open-source. We thank A. Cumming, R. Pinto, C. Zanni and P. Zarka
for inspiring discussions on star-planet magnetized interactions. This
work was supported by the ANR TOUPIES and the ERC project
STARS2. We acknowledge access to supercomputers through GENCI project
1623 and Prace infrastructures. A. Strugarek acknowledges support from
the Canada's Natural Sciences and Engineering Research Council.  

%\bibliographystyle{apj}
%\bibliography{../mybib}

\begin{thebibliography}{32}
\expandafter\ifx\csname natexlab\endcsname\relax\def\natexlab#1{#1}\fi

\bibitem[{Barnes {et~al.}(2011)Barnes, Meadows, Domagal-Goldman, Heller,
  Jackson, L{\'o}pez-Morales, Tanner, Gomez~Perez, \& Ruedas}]{Barnes:2011tc}
Barnes, R., Meadows, V.~S., Domagal-Goldman, S.~D., {et~al.} 2011, 16th
  Cambridge Workshop on Cool Stars, 448, 391

\bibitem[{Cohen {et~al.}(2010)Cohen, Drake, Kashyap, Sokolov, \&
  Gombosi}]{Cohen:2010jm}
Cohen, O., Drake, J.~J., Kashyap, V.~L., Sokolov, I.~V., \& Gombosi, T.~I.
  2010, ApJ, 723, L64

\bibitem[{Cohen {et~al.}(2011)Cohen, Kashyap, Drake, Sokolov, Garraffo, \&
  Gombosi}]{Cohen:2011gg}
Cohen, O., Kashyap, V.~L., Drake, J.~J., {et~al.} 2011, ApJ, 733, 67

\bibitem[{Cuntz {et~al.}(2000)Cuntz, Saar, \& Musielak}]{Cuntz:2000ef}
Cuntz, M., Saar, S.~H., \& Musielak, Z.~E. 2000, ApJ, 533, L151

\bibitem[{Donati {et~al.}(2008)Donati, Moutou, Fares, Bohlender, Catala,
  Deleuil, Shkolnik, Cameron, Jardine, \& Walker}]{Donati:2008hw}
Donati, J.-F., Moutou, C., Fares, R., {et~al.} 2008, MNRAS, 385, 1179

\bibitem[{Fares {et~al.}(2010)Fares, Donati, Moutou, Jardine, Grie{\ss}meier,
  Zarka, Shkolnik, Bohlender, Catala, \& Cameron}]{Fares:2010hq}
Fares, R., Donati, J.-F., Moutou, C., {et~al.} 2010, MNRAS, 406, 409

\bibitem[{Goldreich \& Lynden-Bell(1969)}]{Goldreich:1969kf}
Goldreich, P., \& Lynden-Bell, D. 1969, ApJ, 156, 59

\bibitem[{Ip {et~al.}(2004)Ip, Kopp, \& Hu}]{Ip:2004ba}
Ip, W.-H., Kopp, A., \& Hu, J.-H. 2004, ApJ, 602, L53

\bibitem[{Jardine \& Collier~Cameron(2008)}]{Jardine:2008ec}
Jardine, M., \& Collier~Cameron, A. 2008, A{\&}A, 490, 843

\bibitem[{Kasting {et~al.}(1993)Kasting, Whitmire, \&
  Reynolds}]{Kasting:1993hw}
Kasting, J.~F., Whitmire, D.~P., \& Reynolds, R.~T. 1993, Icarus, 101, 108

\bibitem[{Keppens \& Goedbloed(2000)}]{Keppens:2000ea}
Keppens, R., \& Goedbloed, J.~P. 2000, ApJ, 530, 1036

\bibitem[{Kivelson {et~al.}(2004)Kivelson, Bagenal, Kurth, Neubauer, Paranicas,
  \& Saur}]{Kivelson:2004vf}
Kivelson, M.~G., Bagenal, F., Kurth, W.~S., {et~al.} 2004, In: Jupiter. The
  planet, 513

\bibitem[{Laine \& Lin(2011)}]{Laine:2011jt}
Laine, R.~O., \& Lin, D. N.~C. 2011, ApJ, 745, 2

\bibitem[{Laine {et~al.}(2008)Laine, Lin, \& Dong}]{Laine:2008dx}
Laine, R.~O., Lin, D. N.~C., \& Dong, S. 2008, ApJ, 685, 521

\bibitem[{Lanza(2008)}]{Lanza:2008fn}
Lanza, A.~F. 2008, A{\&}A, 487, 1163

\bibitem[{Lanza(2010)}]{Lanza:2010bo}
---. 2010, A{\&}A, 512, 77

\bibitem[{Lanza(2013)}]{Lanza:2013gj}
---. 2013, A{\&}A, 557, 31

\bibitem[{Lovelace {et~al.}(1986)Lovelace, Mehanian, Mobarry, \&
  Sulkanen}]{Lovelace:1986kd}
Lovelace, R. V.~E., Mehanian, C., Mobarry, C.~M., \& Sulkanen, M.~E. 1986,
  ApJS, 62, 1

\bibitem[{Matt \& Balick(2004)}]{Matt:2004kd}
Matt, S., \& Balick, B. 2004, ApJ, 615, 921

\bibitem[{Matt {et~al.}(2012)Matt, MacGregor, Pinsonneault, \&
  Greene}]{Matt:2012ib}
Matt, S.~P., MacGregor, K.~B., Pinsonneault, M.~H., \& Greene, T.~P. 2012,
  ApJL, 754, L26

\bibitem[{Mignone {et~al.}(2007)Mignone, Bodo, Massaglia, Matsakos, Tesileanu,
  Zanni, \& Ferrari}]{Mignone:2007iw}
Mignone, A., Bodo, G., Massaglia, S., {et~al.} 2007, ApJS, 170, 228

\bibitem[{Miller {et~al.}(2012)Miller, Gallo, Wright, \&
  Dupree}]{Miller:2012gq}
Miller, B.~P., Gallo, E., Wright, J.~T., \& Dupree, A.~K. 2012, ApJ, 754, 137

\bibitem[{Pont(2009)}]{Pont:2009ip}
Pont, F. 2009, MNRAS, 396, 1789

\bibitem[{Russell(1993)}]{Russell:1993jk}
Russell, C.~T. 1993, Reports on Progress in Physics, 56, 687

\bibitem[{Strugarek {et~al.}(2012)Strugarek, Brun, \& Matt}]{Strugarek:2012th}
Strugarek, A., Brun, A.~S., \& Matt, S. 2012, in SF2A-2012: Proceedings of the
  Annual meeting of the French Society of Astronomy and Astrophysics. Eds.: S.
  Boissier, 419--423

\bibitem[{Strugarek {et~al.}(2013)Strugarek, Brun, Matt, \&
  Reville}]{Strugarek:2013uh}
Strugarek, A., Brun, A.~S., Matt, S.~P., \& Reville, V. 2013, In preparation

\bibitem[{Ustyugova {et~al.}(1999)Ustyugova, Koldoba, Romanova, Chechetkin, \&
  Lovelace}]{Ustyugova:1999ig}
Ustyugova, G.~V., Koldoba, A.~V., Romanova, M.~M., Chechetkin, V.~M., \&
  Lovelace, R. V.~E. 1999, ApJ, 516, 221

\bibitem[{Vidotto {et~al.}(2010)Vidotto, Opher, Jatenco-Pereira, \&
  Gombosi}]{Vidotto:2010iv}
Vidotto, A.~A., Opher, M., Jatenco-Pereira, V., \& Gombosi, T.~I. 2010, ApJ,
  720, 1262

\bibitem[{Washimi \& Shibata(1993)}]{Washimi:1993vm}
Washimi, H., \& Shibata, S. 1993, MNRAS, 262, 936

\bibitem[{Weber \& Davis(1967)}]{Weber:1967kx}
Weber, E.~J., \& Davis, L.~J. 1967, ApJS, 148, 217

\bibitem[{Zanni \& Ferreira(2009)}]{Zanni:2009kc}
Zanni, C., \& Ferreira, J. 2009, A{\&}A, 508, 1117

\bibitem[{Zarka(2007)}]{Zarka:2007fo}
Zarka, P. 2007, Planetary and Space Science, 55, 598

\end{thebibliography}

\end{document}